\documentclass[aps,pra,amssymb,nofootinbib]{revtex4}
\usepackage{graphicx}
\begin{document}
\title{Time of arrival with resonances: Beyond scattering states}
\author{Lucas Lamata {\sl and} Juan Le\'on}
\affiliation{Instituto de Matem\'aticas y F\'{\i}sica Fundamental\\
Serrano 113-B, 28006 MADRID}
\date{July 24, 2003}

\begin{abstract}

1$^{st}$ Part: {\bf Assorted questions}.
\begin{itemize}
\item Time as a parameter in Quantum Mechanics.
\item No-Go theorems for a time operator.
\item Localization, time and causality.
\item Causality violation.
\item Localization again.
\end{itemize}
 Lesson 1: {\sl Evading the troubles: Im E$\neq$ 0}.
 
 \vspace*{0.1in}

2$^{nd}$ Part:{\bf Lights and shadows of a time operator}.
\begin{itemize}
\item ``Table-top Spacetime" quantum mechanics.
\item Biphotons at Berkeley.
\item Time operator build-up.
\item Good news - bad news.
\end{itemize}
Lesson 2: {\sl We need the resonances to tell} $\mathbf{1}$.
\end{abstract}

\maketitle

\section*{Assorted questions}
In non-relativistic -- as well as in relativistic quantum mechanics -- 
time $t$ is the parameter that rules time evolution. 
For any operator $A$ in the Hilbert Space:
$$
\frac{d A}{dt}= -i [H,A],\;\; \Rightarrow A(t) $$
The same applies to position $x$: it is the parameter that 
rules spatial displacements:
$$
\frac{d A}{dx}= i [P,A],\;\; \Rightarrow A(x) $$
Coordinate representation comes from the first days of 
quantum mechanics $\psi(x) = \langle x | \psi \rangle$. In 
other words, a position operator $\widehat{position}$ does exist such that
 $\widehat{position} |x\rangle = x |x\rangle$. An immediate consequence is the 
 probabilistic interpretation for the wave function: The probability 
 of finding the particle at $x$ is $P_\psi (x) = 
 |\langle x | \psi \rangle|^2$. In the non-relativistic case the picture
 is more vivid as there $\langle x' | x \rangle = \delta (x'-x)$.
 
 The very existence of $\widehat{position}$ prompts for the question: Does exist a 
 time operator $\widehat{time}$, such that $\widehat{time} |t\rangle = t |t\rangle$?
An affirmative answer would lead us to consider  $P_\psi (t) = 
 |\langle t | \psi \rangle|^2$ as the probability of finding the particle at the instant
 t. Needless to say, the temptation of understanding ``finding" in terms of ``running into"
 is difficult to resist.
 
 In his contribution to the Encyclopedia of Physics~\cite{pauli}, Pauli advised against a time operator.
 His argument can be understood in terms of the Stone-Von Neumann theorem~\cite{neumann}: 
 Given two self adjoint operators $\hat{A}, \hat{B}$:
 $$ 
 [\hat{A}, \hat{B}]=i \Leftrightarrow \sigma(\hat{A})= \sigma(\hat{B})$$
 where $\sigma$ denotes the operator spectrum. Now, for elementary systems $\sigma(\hat{t})$ is 
 the real line, while $\sigma(\hat{H})\geq m$, where $m$ is the particle mass. Therefore, if the system 
 has a Hamiltonian either: There is no time operator conjugate to $\hat{H}$, or $\hat{t}$ is not self-adjoint.
 
 The problem appears in the Poincare group once a mass shell (a representation of the 
 elementary system) is chosen. In fact, if $\hat{P}^2 = m^2$ and there exist a conjugate
 position operator $\hat{X}$, then operating with $\hat{X}$ throws the particle momentum out of the
 mass shell:
 $$
 e^{-i \hat{X} p' } \hat{P} e^{i \hat{X} p' }= \hat{P} + p' ,\,\, 
 \mbox{but} (\hat{P} + p')^2 \neq m^2 $$
 This is among the reasons for demote time and position to mere parameters while promoting the fields 
 to operators in quantum field theory.
 
 In 1974 Hegerfeldt started~\cite{hegerfeldta} a far reaching line of research on the localization of elementary systems that concluded 
 with very strong results. In poor words we could say that: \begin{center} "Strict localization of a system whose Hamiltonian is bounded from
 below is incompatible with causality."\end{center}  
 Technically speaking: with $ \hat{H}\geq c$ and $\psi_t =e^{-i H t}
  \psi_0$,  one can define $\forall$ self adjoint positive operator $\hat{A}\geq 0$ the bracket 
  ${\mathcal P}_A(t)=\langle \psi_t|A|\psi_t \rangle$. Hegerfeldt shows that:  
\begin{center} {\sl Either} ${\mathcal P}_A  (t)\neq 0\,\forall \, t 
  \in \mathbf{R} $ {\sl or} ${\mathcal P}_A(t)= 0\,\forall \, t 
  \in \mathbf{R}$.\end{center} The relation with localization appears by choosing
  $A=\int_V |x\rangle \langle x|$ with $V$ a Borel set in $\mathbf{R}^3$. Then: 
\begin{center} {\sl Either} $\psi $ is in $V$ forever
 $ (P_V (t)\neq 0 \,\forall \, t 
  \in \mathbf{R} $) {\sl or} $\psi $ is never in $V$  
  $(P_V (t)= 0 \,\forall \, t 
  \in \mathbf{R})$.\end{center} 
  
Positivity of energy implies also causality violation, at least in the form of superluminality. Consider spontaneous emission from an initially excited atom A$^*\rightarrow$ A $+\gamma$:
\vspace{0.3in}
\begin{center}
\begin{tabular}[t]{lcl}
 Excited atom $|\psi\rangle$\hspace{1in} & No photons& \hspace{1in}Detector $|D_0\rangle$\\
&.&  \\
&.&\\
&.&\\
 Final state atom $|\psi_t\rangle$& n photons&\hspace{1in} Excited detector $|D_e\rangle$\\
\end{tabular}
\end{center}
When the detector is excited by the photon it clicks with probability,
$$ {\mathcal P}_{\mbox{click}}(t)
=\langle \psi_t|{\mathcal O}_{\mbox{click}}|\psi_t\rangle$$

The operator ${\mathcal O}_{\mbox{click}}$ projects over all excited states of the detector: ${\mathcal O}_{\mbox{click}}= 
\sum_e |D_e\rangle \langle D_e|$ therefore according to Hegerfeldt:
\begin{center} {\sl Either} ${\mathcal P}_{\mbox{click}}(t)=0 \,\forall \, t$ {\sl or} ${\mathcal P}_{\mbox{click}}(t)>0 \,\forall \, t$\footnote{N.B. For good mathematical reasons, the condition $\forall \, t$ above should be understood as ``for almost every $t$".}. \end{center}
This could be summarized according to the graphics below:


\begin{figure}[h] 
 \begin{center}
 \includegraphics{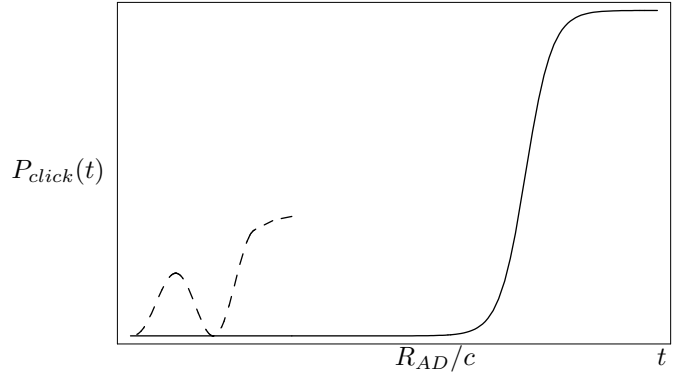}
 \end{center}
 \caption{Solid line: Causal expectation for the time behaviour of the probability of the first click. Dashed line: generic time behaviour of the probability of the first click as explained in the text. } \label{Fig.1}
\end{figure}
where the dotted curve is a generic example of what may arise from the analysis of Hegerfeld, while the solid one is what comes out 
from causality expectation ($t_{\mbox{click}} \geq R_{AD}/c$).
 This remarkable result led Prigogine and coworkers to analyze it in more detail~\cite{prigogine}, concluding that positivity of energy was incompatible with simultaneous strict localization of the wave function and of its first time derivative. Then, with locality lost, there is no real causality violation, only the appearance of it. In more terrestrial words, the apparently localized wave is the superposition of two opposite moving waves that spread to infinity, whose amplitudes interfere destructively, such that at $t=0$ their sum vanishes  everywhere except in the ``localization volume".

The localization of particles can not go beyond the limit imposed by the Paley-Wiener~\cite{paley} Theorem XII. Consider  two Fourier related functions $f(t)$ and $\tilde{f}(k)$,  that are square integrable $f,\tilde{f} \in L^2(-\infty, \infty)$. Then, there is a general relation between the support of one of these functions (assumed to be bounded) and the asymptotic behaviour of the other. This is depicted in Figure 2.

\begin{figure}[h]
 \begin{center}
 \includegraphics[width=15cm]{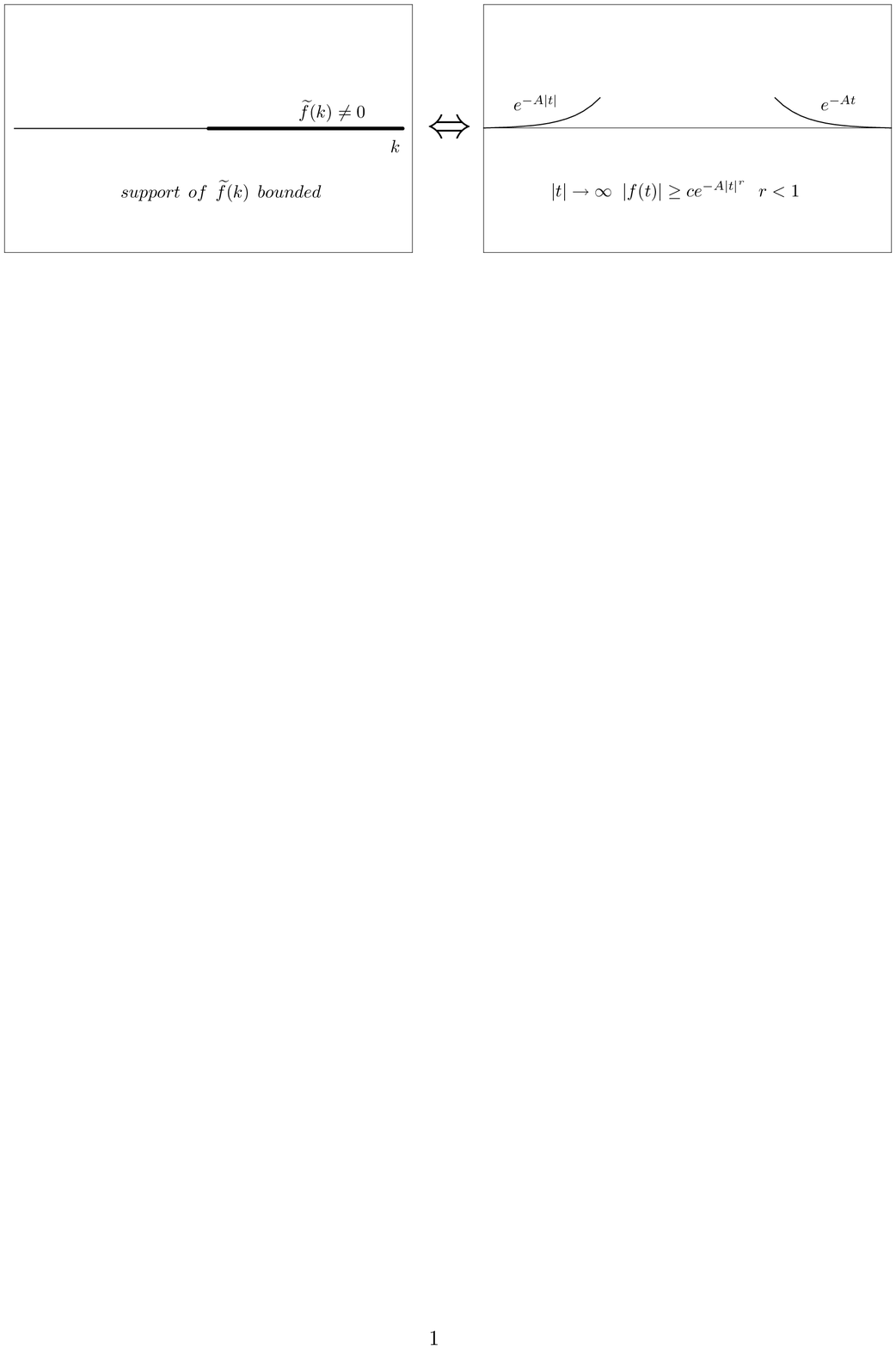}
 \end{center}
 \caption{The Paley-Wiener theorem for two Fourier related square integrable functions $f$ and $\hat{f}$. } \label{Fig.2}
\end{figure}

I. Bialynicki-Birula applied this theorem to one photon states getting~\cite{iwoc} the best location attainable for these states.  The recipe is the following: i) Define the one foton mode $\tilde{f}(\mathbf{k},\lambda)$ as:
$$ |1_f\rangle = \sum_\lambda \int d^3 k \tilde{f}(\mathbf{k},\lambda) a^\dag (\mathbf{k},\lambda)|0\rangle$$
ii) Take into account the probabilistic nature of the wave function (or $\tilde{f}\in L^2$)
$$\sum_\lambda \int d^3 k \,\,|\tilde{f}(\mathbf{k},\lambda)|^2=1$$
and iii) The fact that to represent a photon, $\tilde{f}(\mathbf{k},\lambda))=0$ for $\omega_k <0$. Then it is possible to conclude from the theorem that (written here in one dimensional notation): 
$$f(x,t)= Z(t+x/c)+Z(t-x/c)$$
$Z$ vanishes slowlier than any exponential:
$$\lim_{|\tau|\rightarrow \infty} Z(\tau)\,\,{{\gtrsim} } \,\, e^{-A \tau^\gamma},\,\,\, A>0, \gamma <1$$
This was the best photon localization theoretically attainable in 1998.

The quest for photon localization moved forward recently due to the analysis made in Ref.~\cite{eberly}  of the  final state entanglement between the emitted  photon and the final atom produced by spontaneous decay of a simple two level atom. The atom, initially in the excited state $|e\rangle$, gains recoil momentum $\mathbf{p}'$ by emitting a photon of momentum $ \mathbf{k}$ while dropping into its ground state $|g\rangle$.
\begin{figure}[h]
  \vspace{-1cm}\begin{center}\includegraphics[width=10cm,height=7cm]{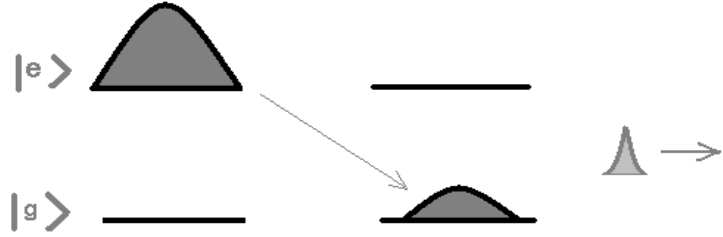}
  \end{center}
\label{Fig.3}\caption{Spontaneous emission of a photon from a
two-level atom.}
 \end{figure}
The Hamiltonian of the system is the sum of the centre of mass Hamiltonian, the e.m. field Hamiltonian and the dipole coupling between the atom and the field. The initial state of the atom can be written in the form 
$$|\psi(0)\rangle =\int d^3 p  \,\, \exp{(-{\mathbf{p}}/ \,\Delta p)^2}|e; {\mathbf{p}}\rangle \otimes |0\rangle$$
where  a Gaussian distribution for motional momenta is allowed to the atom and $|0\rangle$ stands for no photons. After emission the state is:
$$|\psi(t)\rangle =\int d^3 p' d^3 k \,\,  \Psi({\mathbf{p}}',{\mathbf{k}})\,e^{-i(E_{p'}+k c)t}\,
|g; \mathbf{p}'\rangle \otimes |1_{\mathbf{k}}\rangle $$
This expression assumes that $t>>1/\gamma$, where $\gamma$ is the radiative linewidth. $\Psi(\mathbf{p}' ,\mathbf{k})$ is the probability amplitude for finding the photon and the atom with these final momenta. For a realistic atomic system it is possible to write in the case of opposite final momenta:
$$\Psi(p',k) \propto \frac{e^{-(\delta p'/\Delta \omega)^2}}{\delta p' +\delta k+i \gamma}$$
The objects in this expresion are\footnote{Notice that we are using $\hbar=1, c=1$ in the paper, so that these quantities appear only where they serve as sighting aids}  $\gamma$, $\delta k=k-\omega_0$ and $\delta p'=(p'-\omega_0) (\hbar \omega_0/mc^2) $ where $m$ is the atom mass and $\hbar \omega_0$  the $e-g$ levels energy difference. Finally, $\Delta \omega$ is the motional Doppler width

As noted by the authors, the Lorentzian factor in the above expression for $\Psi$ is responsible for the entanglement between the photon and the atom, while the Gaussian gives the range of relevant atom momenta. The authors then decompose the wave function $\Psi$ into a sum of mode products.
$$ \Psi(p',k)=\sum_n \lambda_n \psi_n(p') \phi_n(k)$$
 Now, the localization of the photon is ruled by the entanglement instead of by the photon wavelength as naively expected. More remarkably, the photon modes $\tilde{\phi}_n (x,t)$ have Gaussian \nolinebreak[4] tails! The next figure shows the modes  with $n=1$ for $t=5/\gamma$
\begin{figure}[h] 
 \begin{center}
 \includegraphics{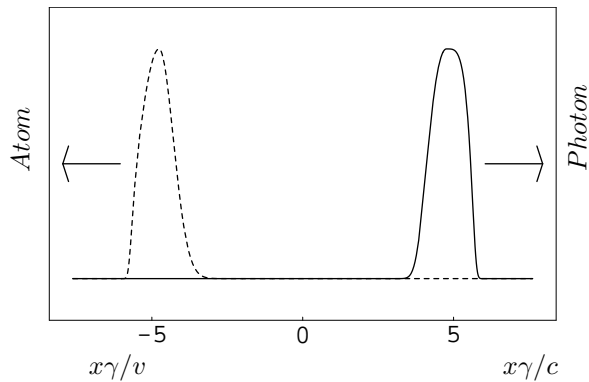}
 \end{center}
 \caption{Atom and photon wave packets flying appart from each other at speeds v and c at $t=5\,\gamma^{-1}$. Notice that the packets are centered at
 x$\gamma/$v=-5 and x$\gamma/$c= 5 respectively.} \label{Fig.4}
\end{figure}

The figure shows clearly that, instead of the result depicted in Fig. 1, the photon mode presents a nice sharp edge at $x/c=5/\gamma$, corresponding to the initiation of decay at $t_0=0$. This, along with the Gaussian tails, is valid for the other modes and other values of $t$. The alert reader has probably noticed already how singular the limit $\gamma \rightarrow 0$ is. In fact, for $\gamma$ finite, the spectrum of the Hamiltonian is not contained in the real line, therefore conditions like $H \geq c$ loose much of its meaning. 

\section*{Lights and shadows of a time operator}

Look back for a moment to two kinds of traditional experiments of the last century: namely ``beam-target" and ``beam-beam" collisions. The need of a sound theoretical foundation to analyze their results produced the $\mathcal{S}$ matrix.
The goal was that initial states, prepared at $t=-\infty$ (!), and final states, to be detected at $t=+\infty$ (!), could be mathematically related by the $\mathcal{S}$ matrix, ${\mathcal{S}} (E,\Omega)$, depending of the energy $E$ and the scattering angle $\Omega$. The point of interest to us here is the inescapable ``infiniteness" imposed by this theoretical frame on the experimental conditions: infinite lapse, infinite distance, stationary conditions $\dots$ The fast pace of technical progress has made possible to turn beams into single particles, incident one by one, and to shrink the sizes of experimental scrutiny, to the microns, angstroms or femtoseconds. Needless to say, in these conditions we need, going beyond the standard $\mathcal{S}$ matrix, of a theoretical tool capable to deal with finite (short, almost infinitesimal) lapses and ranges.
\begin{figure}[h]
 \begin{center}
\includegraphics[width=10cm]{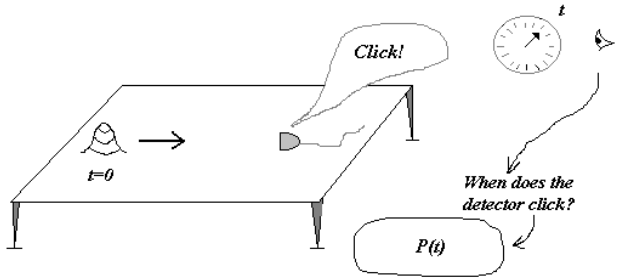}
 \end{center}
 \caption{Idealized experimental setup for the measurment of the time of arrival at the detector}
\label{Fig. 6}
\end{figure}

Current experiments could be included in a category of desktop setups in which a system is prepared at $t=0$, evolves in some tricky way (that we want to understand) during a finite time and, finally, produces (or stops from producing) a click on a nearby detector. The observer is watching the instant when the detector clicks and records the time elapsed. This becomes a central question: When the detector clicks? So, theory and experiment gather now around a probability $P(t)$ in the time domain.\footnote{Of course, ``when" and ``where" could be traded - one by the other - depending of the experiment.} This is far away from the infinite lapse proper of the $\mathcal{S}$ matrix.  

A nice example of the above situation is the series of experiments~\cite{berkeley1,berkeley2} carried out at Berkeley during the first half of the last decade. In very simplified terms, assume a continuous laser beam that goes through a nonlinear KDP crystal. Filter only those photon pairs produced by parametric down conversion with identical frequencies. Send both, idler and signal photons, by different paths to interfere at a Hong-Ou-Mandel interferometer~\cite{hong}. Put a detector at each of the interferometer exit ports and record the rate of coincidence detections.
\begin{figure}[h]
\begin{center}
\includegraphics{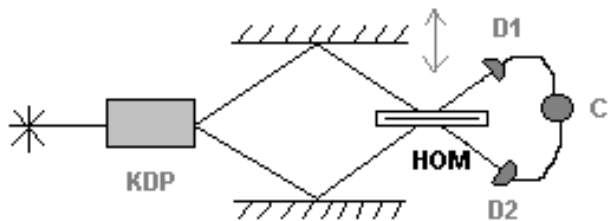}
\end{center}
\caption{Berkeley biphoton experiment. KDP is the nonlinear
crystal. HOM is the Hong-Ou-Mandel beam splitter. D1 and D2 are
detectors and C is the coincidence counter.}
\end{figure}

Both, idler and signal photons, can be transmitted, can be reflected, or one transmitted and the other reflected. 
In the first two cases they produce coincidence counts, giving a count in the coincidence detector. However, if they arrive simultaneously (within their coherence time), they interfere destructively (this is the reason for putting the HOM beam splitter) and there is no coincidence count. In the experiment, one of the mirrors can be displaced using an inchworm, modifying the difference $\delta$ in the optical paths of the two photons. The rate of coincidence counts for each $\delta$ is recorded and a minimum is observed at $\delta=0$. It is remarkable that, while the detectors can not separate below $\mu$seconds, the width at half depth of the interferometer dip is of $20$ fs approximately.
\begin{figure}[h]
 \begin{center}
\includegraphics{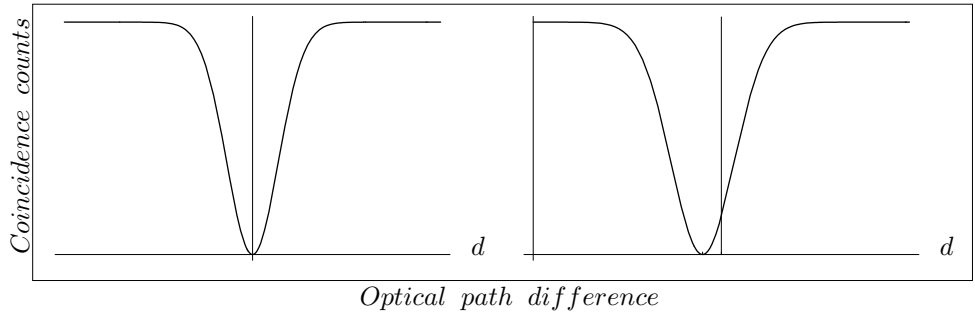}
 \end{center}
 \caption{Coincidence counts vs. optical path difference. Left: Free paths. Right: A photonic band gap (HL)$^n$H intercepts the signal photon path, the minimum is now at $-\delta/c$ fs. Both dips have similar widths due to the fact that idler and signal photons are entangled in a  biphoton state}
\label{Fig. 6}
\end{figure}
The experiment was repeated putting a photonic band gap\footnote{The notation H (L) stands for the consecutive high (low) refractive index layers of the PBG} (PBG) in the path of the signal photon. That produced an effective shortening in the optical path of this photon and hence a negative time delay $-\delta/c$. The group velocity 
in the PBG was measured to be $u\approx 1.6 c$. After the wealth of controversy~\cite{heitmann,garrison} consequence of this result, there is a general agreement that
causality is not violated by these faster-than-light-photons. 

We now turn to the specific question of  whether in quantum mechanics we can tell (and how)   the instant in which a system acquires a certain property of our interest. The simplest of these is the time of arrival at -or of pass by- a certain point of space. In classical dynamics there is a well known answer:
$$t({\mathbf{x}})=\int_{\mathbf{q}}^{\mathbf{x}}\frac{m \,d{\mathbf{q}'}}{\sqrt{2 m(H({\mathbf{q}}',{\mathbf{p}}')-V({\mathbf{q}}'))}}$$
But notice that this apparent simplicity is deceptive: for non-integrable systems $t(\mathbf{x})$ may be path dependent.
For the free particle the above gives $t(x)=m (x-q)/p$ where $q$ and $p$ are the initial position and momenta respectively. Below, we shall give explicitly the quantum version of this object, that has been a source of abundant literature~\cite{muga}. For the time being, we assume that it is a known operator $\hat{t}_0$, conjugate to the free Hamiltonian $\hat{H}_0$, that is, such that $[\hat{H}_0,\hat{t}_0]=i \hbar$. We now use the quantum canonical transformation $\Omega$  transforming the free Hamiltonian into the complete one  $\hat{H}=\hat{H}_0 + V$, with the idea that the same transformation shall turn $\hat{t}_0$ into the  time of arrival operator $\hat{t} $ appropriate in the presence of interactions:
\begin{center}
\begin{tabular}{ccc}
 $\hat{H}_0$&$---\longrightarrow$ & $\hat{H}=\Omega \,\,\hat{H}_0\,\,\Omega^{-1}$\\
 &&\\
&&\\
  $\hat{t}_0$&$---\longrightarrow$ & $\hat{t}=\Omega \,\,\hat{t}_0\,\,\Omega^{-1}$ \\
\end{tabular}
 \end{center}
 Notice that this procedure guaranties that $\hat{t}$ is an  operator conjugate to $\hat{H }$, a basic requirement for a sound time of arrival operator.

Classically $\Omega^{-1}$ is the Jacobi-Lie  transformation to action-angle variables. Quantum mechanically, $\Omega$ is the M\"oller wave operator~\cite{moller} that connects the eigenstates of $\hat{H}_0$ to the actual energy eigenstates:
$$
\begin{array}{lcr}
\begin{array}{rcl}
\hat{H}_0\,\phi^0_E&=&\, E\,\phi^0_E\\
&&\\
\hat{H}\,\phi_E&=&\, E\,\phi_E\\
\end{array}&\Huge\}\,\,\Rightarrow \hspace{1cm}&
\Omega=\sum\int_{\sigma(E)} \,dE\,\,|\phi_E\rangle\,\langle \phi^0_E|\\
\end{array}$$
We now proceed in parallel with the eigenstates of the time operator~\cite{leonb}:

$$
\begin{array}{lcr}
\begin{array}{rcl}
\hat{t}_0\,\varphi^0_t&=&\, t\,\varphi^0_t\\
&&\\
\hat{t}\,\,\,\varphi_t&=&\, t\,\varphi_t\\
\end{array}& \Huge\}\,\,\Rightarrow\hspace{1cm}&
\begin{array}{ccc}
|\varphi_t\rangle&=& \Omega\,\, |\varphi_t^0\rangle\\
&&\\
\hat{t}&=& \Omega \,\,\hat{t}_0\,\,\Omega^{-1}\\
\end{array}
\end{array}$$

We showed above the very simple expression for the classical time of arrival of the free particle $t_0(x)\,=\,m (x-q)/p$.
A properly ordered quantum version is 
$$\hat{t}_0(x)\,\,=\,\,-\, e^{-i \hat{p}x}\,\,\sqrt{\frac{m}{\hat{p}}}\,\, \hat{q}\,\, \sqrt{\frac{m}{\hat{p}}}\,\,e^{i \hat{p}x}$$
Notice that $\hat{p},\hat{q}$ are operators while $x$ is a parameter and $m$ the particle mass. Here, the roles of space and time are the opposite than in the standard formulation: Here $t_0$ changes as $x$ changes. In other words, the instant of arrival depends parametrically on the arrival position. 

All the steps leading to the quantities of interest  in the presence of a potential $\hat{H}=\hat{H}_0 + V$ have been spelled~\cite{leonb} in detail in the literature. The final result is that the eigenstates for the arrival at $x$ in the instant $t$ are 
$$|t\,x\rangle=\, (\frac{2\hat{H}}{m})^{1/4} \,e^{-i\,\hat{H}\,t}\,\,|x\rangle $$

This is a highly symbolic  notation because $x$, the detector position, is not a property of the particle, but something alien to it. Accepting this  as it is, the result is a time eigenstate in terms a position (the detector location) eigenstate. The spectral decomposition of the time of arrival operator is given, in terms of these eigenstates as:
$$ \hat{t}(x)\,=\, \int dt \,\,t\,\,|t\,x\rangle\,\langle t\,x|$$
Following the Born rule, the probability that a particle prepared in the state $\psi$ -at $t=0$, as everything here is in the Heisenberg picture- arrives at $x$ in the instant $t$ is $|\,\langle t\,x|\,\psi\rangle\,|^2$. There are some subtleties in the more realistic case of 3 space dimensions~\cite{leona}. In particular not all the states that one can prepare at $t=0$ may be eventually detected. On the contrary, the detected states form a subspace of the Hilbert space, and it is necessary to project on it to get physically meaningful predictions.

Armed with this artillery, we attack again the questions of physical interest that we abandoned paragraphs above. In particular, if we put a potential (a barrier, a well, a dielectric, a force field, etc) in between the initial state and the detector, What should we expect for the distribution of times of arrival? That is: We record the time of arrival 
of the particle at the detector, then repeat the experiment once and again with the same initial state and record the successive  times of arrival attained. At the end of the day, the experiment has accumulated  a statistical distribution in times of arrival to be compared with the theoretical predictions:~\cite{leonb} $|\langle t\,x|\,\psi\rangle|^2 dt$, for the probability density, $\langle \psi\,|\,t(x)\,|\,\psi\rangle$ for the average time of arrival, etc. The obvious question is: Do they match?

We have devoted some time to this question and the answer is a mixture of yes and noes. The formalism is a step forward with respect to other tunneling times that appear in the literature in the sense that it also predicts the probability densities in times of arrival to be recorded experimentally. This goes far beyond the Wigner time delay and other tunneling times, that are just a number for each experiment. In fact, our result for the time of arrival can be read as the quantum average of the Wigner time delay, but in addition predicts probabilities, r.m.s. uncertainties, and the like.

Long time ago, Hartman~\cite{hartman} explored the times involved in the crossing of classically forbidden barriers, reaching the conclusion that the crossing was instantaneous. He explained the effect using heuristic arguments in terms of the Wigner time delay. The time of arrival formalism gives an explanation using the basic postulates of quantum mechanics. 

We have applied the formalism~\cite{lucas2} to the biphoton Berkeley experiments. It is possible to get~\cite{leond} the photon time of arrival, in vacuum and in a passive dielectric. Therefore one can predict the distribution in times of arrival of both components in the biphoton, the idler and the signal and compare with the experiment. Experiment and theory match~\cite{lucas2}. Note that  these experiments are just exercises in detecting times of arrival. It is not surprising that in order to describe the experiment, the standard theoretical treatment (in which operators or wave functions are functions of time), has to be taylored to finally match the time of arrival formalism.

We  also explored other physical systems, in particular those in which  resonances play a prominent role. In spite of our earlier  expectations, the formalism was not able to handle these situations.  On the contrary, as a predictor of the time of arrival in the presence of resonances, it was a complete failure.
In the rest of the paper we present in simple terms the failure and the lesson that we extracted from it. 

Consider a state $\psi$ prepared at the left of a potential region and place an (ideal) detector at a position $x$ at the right of that region. The amplitude for detecting the particle at $x$ at the time $t$ is:
$$\langle t x| \psi\rangle \,=\, \langle  x|e^{-i \hat{H}t}\, \left(\frac{ 2\hat{H}}{m }\right)^{1/4}\, |\psi\rangle
\,=\,\int dE \,\,e^{-i Et}\, \left(\frac{ 2E}{m }\right)^{1/4}\,
\langle  x|E\rangle\,\,\langle E|\psi\rangle$$
In the above expression we have introduced a decomposition of the identity in the form $\int\,dE \,\,|E\rangle\,\,\langle E|$. This is a scattering situation, therefore with $\psi$ prepared at the left of the barrier $\langle E|\psi\rangle \simeq \hat{\psi}(p)$, where $\hat{\psi}(p)$ is the initial wave packet in momentum space. Also, with the detector placed at the right of the barrier $$\langle x|E\rangle \simeq T(p) e^{i px}$$
Now everything is clear: we get no trace of resonances because we only used scattering states.  Perhaps, we should get a more complete answer if we plug into the formalism a richer decomposition of the unity. More specifically, Would be possible to predict the delays due to the existence of resonances?. Yes, but incorporating Gamow states into the space of intermediate states.

ACKNOWLEDGMENTS

J. L. is indebted to Lidia Ferreira for invitation to attend the meeting. He also thanks the participants for vivid and engaging discussions and for teaching him a lot about resonances. This work was financed in part by the Spanish Ministry of Science and Technology under contract BFM 2001- 2088.

\end{document}